\documentclass[12pt,preprint]{aastex}
\received{ }
\accepted{ }
\bibpunct{(}{)}{;}{a}{}{,}

\shorttitle{PuMa II}
\shortauthors{Karuppusamy et al}

\begin{document}

\title{PuMa II: A Wide Band Pulsar Machine for the WSRT}
\author{Ramesh Karuppusamy}
\affil{ASTRON, 7990 AA Dwingeloo, The Netherlands}
\affil{Sterrenkunde Instituut, University of Amsterdam}
\affil{Kruislaan 403, Amsterdam, The Netherlands}
\email{ramesh@astron.nl}
\author{Ben Stappers}
\affil{ASTRON, 7990 AA Dwingeloo, The Netherlands}
\affil{Sterrenkunde Instituut, University of Amsterdam}
\affil{Kruislaan 403, Amsterdam, The Netherlands}
\and
\author{Willem van Straten}
\affil{Centre for Astrophysics and Supercomputing}
\affil{Swinburne University of Technology, Melbourne, Australia}

\begin{abstract}

     The Pulsar Machine II (PuMa II) is the new flexible pulsar
 processing backend system at the Westerbork Synthesis Radio Telescope
 (WSRT), specifically designed to take advantage of the upgraded
 WSRT. The instrument is based on a computer cluster running the Linux
 operating system, with minimal custom hardware. A maximum of 160 MHz
 analogue bandwidth sampled as 8$\times$20 MHz subbands with 8-bit
 resolution can be recorded on disks attached to separate computer
 nodes. Processing of the data is done in the additional 32-nodes
 allowing near real time coherent dedispersion for most pulsars
 observed at the WSRT. This has doubled the bandwidth for pulsar
 observations in general, and has enabled the use of coherent
 dedispersion over a bandwidth eight times larger than was previously
 possible at the WSRT. PuMa II is one of the widest bandwidth coherent
 dedispersion machines currently in use and has a maximum time resolution
 of 50ns. The system is now routinely used for high precision pulsar
 timing studies, polarization studies, single pulse work and a variety
 of other observational work.

\end{abstract}


\section{Introduction}

  Pulsars are rotating neutron stars and relatively weak radio
  sources. Observing pulsars requires a large telescope collecting
  area and wide observation bandwidths in order to improve
  sensitivity. Some pulsars have a small rotational period, with the
  fastest spinning at 716 Hz \citep{hrsf+06} and the pulsed emission
  from these sources is often only a small fraction of the pulse
  period, with a duty cycle of 5--10\%. Owing to this nature of
  pulsars, observing them dictates high time resolution and wide
  bandwidths as important requirements in pulsar instrument
  designs. The recent discovery by \citet{he07} of the occurrence of
  bright, extremely narrow pulses of 0.4 ns duration in the Crab
  pulsar further illustrates the importance of high time
  resolution. High time resolution is also a requirement of
  high-precision pulsar timing, an experimental technique with the
  potential for detecting the gravitational wave background using an
  array of millisecond pulsars \citep{srtr90}.

  As the radio signals from pulsars propagate through the interstellar
  medium (ISM), they undergo dispersion, giving rise to the smearing
  of signals across the observed spectral band and limiting the final
  time resolution. If not corrected, the effect of dispersive smearing
  can be severe enough to wipe out the pulsed signal. Dispersion can
  be corrected before or after detection of the pulsar signal. In the
  post-detection method, also called incoherent dedispersion, the
  pulsar signal is first split into narrow channels and the signal in
  each channel is "detected", i.e the voltage is squared to form the
  instantaneous source intensity. The detected time series are then
  shifted in time with respect to a reference channel, and added to
  give an average pulse profile with a high signal-to-noise ratio
  (S/N). However, this method suffers from the disadvantage that the
  residual dispersion smearing is still present in the narrow
  channels, effectively limiting the maximum time resolution
  attainable.

  A better method in which the dispersion is corrected completely is
  called coherent dedispersion; it results in a very high time
  resolution, limited only by the sampling interval, and better
  S/N. This technique, pioneered by \citet{han71}, involves sampling
  and recording raw voltages and then using digital computers to
  invert the effect of the interstellar dispersion on the pulsar
  signals. In practice, this is done by modeling the ISM as a filter
  that imparts a frequency dependent delay on the signal, and
  convolving the data with the inverse of the transfer function of the
  filter.

  Even though the second method removes dispersion completely, it is
  computationally very intensive. Therefore, incoherent dedispersion
  was the technique of choice in early pulsar instruments based on
  mostly analogue filter banks \citep{skn+92}. Other instruments using
  incoherent dedispersion were based on Acoustooptic spectrometers
  \citep{his95} and autocorrelation spectrometers
  \citep{nav94}. Coherent dedispersion in hardware was implemented,
  although these systems had limited bandwidth, with a maximum of 2
  MHz \citep{hsr87}. In the next generation of machines, a combination
  of digital filter banks and hardware coherent dedispersion in the
  narrow channels was implemented \citep{bdz+97} allowing high
  resolution observations with up to 64 MHz bandwidth. Advances in
  digital signal processors made digital filter banks possible
  \citep{vkv02} allowing even wider bandwidths to be used (up to 80
  MHz) and giving a new level of flexibility in terms of a variable
  number of filter channels and limited baseband recording (up to 20
  MHz, 2-bit data).

  Wide bandwidth coherent dedispersion had to wait till technological
  developments in disk and tape based storage in the late-90's that
  made base band recording more accessible
  \citep{jcpu97,sst+00,hot05}. With the improvements in storage
  technology, an even larger bandwidth can now be baseband recorded,
  and relatively cheap cluster computers can be used to coherently
  dedisperse the data in software. The design of Pulsar Machine II
  (PuMa II) has taken advantage of this progress in hardware
  technology. We describe PuMa II in the following sections. The rest
  of the article is organized beginning with section 2 describing the
  WSRT's interface to pulsar instruments. In section 3 the PuMa II
  hardware design, implementation details and software are
  discussed. Some results illustrating the instrument's capabilities
  are presented in section 4. A comparison with other pulsar
  instruments is made in section 5 and conclusions are presented in
  section 6.

  \section{The WSRT Tied Array Interface}
  
  The WSRT is a synthesis array telescope \citep{bh74}. The telescope
  now consists of 14 25-m diameter parabolic telescopes on a 2.7 km
  east-west line. The first ten telescopes are spaced evenly at 144m
  and the last four are movable on rails. For synthesis observations,
  the movable telescopes can be positioned at various locations on
  the rails allowing a favorable $uv$ coverage. The recent upgrade of
  the telescope has resulted in larger bandwidth, frequency agility,
  better telescope surface, newer hardware and software, making the
  WSRT a very sensitive and flexible telescope. The frontend receivers
  in each telescope, called MFFE (Multi Frequency Front End) covers
  frequencies from 110 MHz to 9 GHz in both polarizations almost
  continuously over eight frequency bands. With the highly flexible
  MFFE design, switching to any of the supported frequency bands can
  be done within a minute. The array is most sensitive in the 21cm
  Band with a system temperature of 27K and a telescope gain of 1.2
  K/Jy.
  
  Pulsar observations in synthesis telescopes are less straight
  forward when compared to single dish telescopes. To improve
  sensitivity to pulsars, the signal from all telescopes in the
  synthesis array should be added after the signal from each telescope
  is delayed appropriately. In the WSRT this is done by tapping off
  the digitized signal sent to the correlator and adding the signal
  digitally in the Tied Array Adder Module (TAAM). The TAAM provides
  the added signal as a digital or analogue output to other backend
  systems like pulsar machines and VLBI recorders.
  
\begin{figure}[htbp]
  \epsscale{0.98}
  \plotone{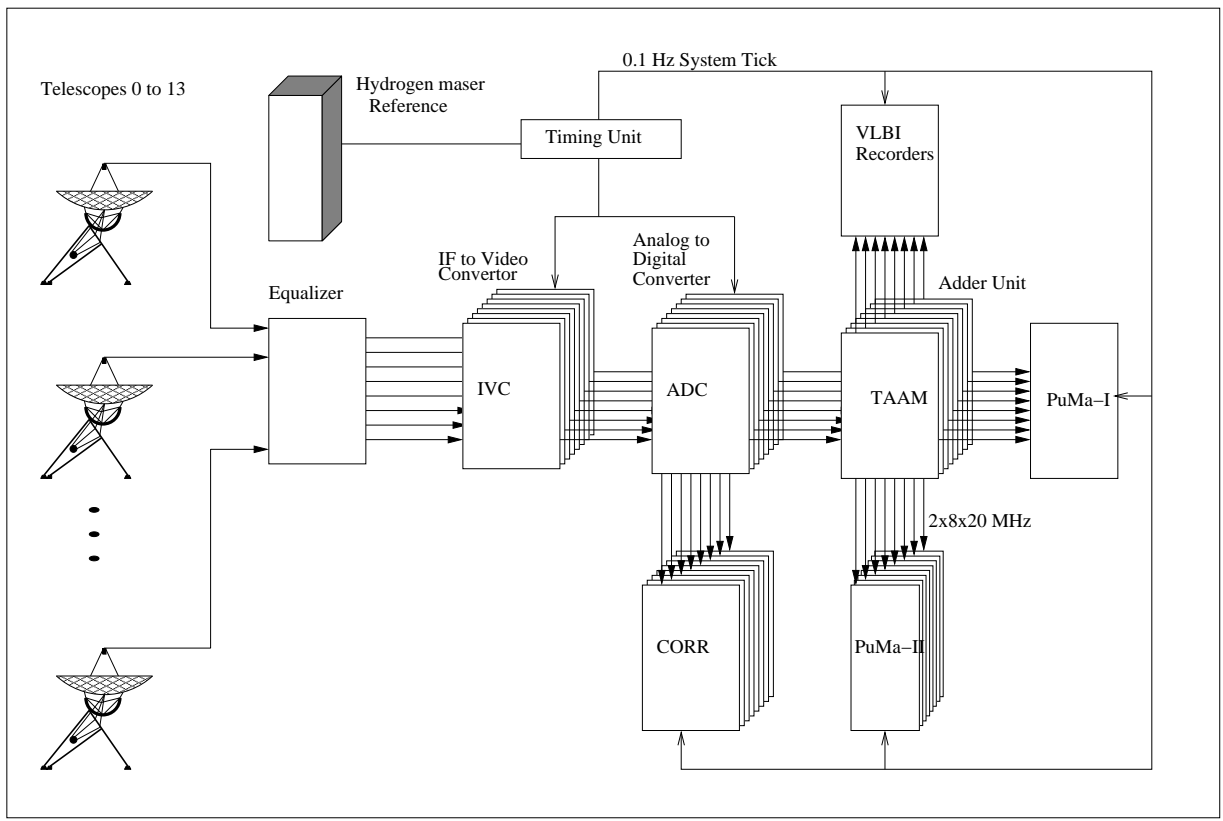}
  \caption{ A block diagram of the WSRT. The radio telescopes are
    equipped with the multi-frequency frontend, MFFE. Pulsar machines
    (includes PuMa II) and other backend instruments requiring the
    tied array outputs are connected to the Tied Array Adding Module
    (TAAM). The correlator system computes all baselines in the
    array. All systems are synchronized by the Hydrogen maser
    reference.}
  \label{fig1}
\end{figure}

  An overview of the WSRT is shown in Fig. \ref{fig1}. The voltages
  induced at the probes corresponding to the two orthogonal
  polarizations in the frontends are down converted to intermediate
  frequency (IF) of 100 MHz at the telescope frontends. All systems
  following this stage are equipped to handle these two orthogonal
  polarization signals. The signal from the frontends are transported
  in phase compensated coaxial cables to the receiver room. The IF
  signal width is 80 MHz (for $F_{sky} < 1$ GHz) or 160 MHz ( $F_{sky}
  > 1$ GHz) depending on the band of operation. The signal then passes
  through an equalizer to compensate for cable losses. The IF to Video
  frequency converter (IVC), splits the IF signals into 20MHz-wide
  subbands and frequency translates to 20 MHz baseband signals. The
  signals are then real-sampled at the Nyquist rate in the analogue to
  digital converter (ADC) units using a 40 MHz clock and two bit
  digitizers. In comparison to the complex sampling used in similar
  instruments, real sampling of the signal results in better quality
  due to the non-ideal realizations of the physical 90$^\circ$
  phase-shifters required in sampling the quadrature signal. A
  geometric delay to the digitized basebands are now applied and are
  sent to both the correlator system and to the adder module,
  TAAM. The TAAM forms the coherent sum of the signals from the
  telescopes in the array. From this point onwards, the data
  distribution is done via optical fibers. The use of optical links
  improves noise immunity in the data communication paths of the
  backend systems by isolating electrical grounds between the
  subsystems. This avoids the formation of electrical ground loops in
  the system. The net effect of the use of fiber links is the
  improvement in signal integrity, thereby improving the overall
  system quality greatly. The TAAM adds the 2-bit data from all 14
  telescopes in phase, resulting in a signal of equivalent strength to
  that from a single dish telescope of 93-m diameter. Addition of the
  2-bit signals from all 14 telescopes results in a 6-bit value. The
  two 6-bit values corresponding to the two polarizations are packed
  as two 8-bit signed integers and sent to the storage nodes in the
  PuMa II cluster via optic fibers.

     The addition of all telescopes signals results in a fan beam,
  with the beam width depending on the wavelength of operation; at 21cm
  this is $28\arcmin \times 0.3\arcmin$ . The narrow beam formed by
  the phased array has some advantages when compared to the beam of an
  equivalent single dish telescope. The synthesized beam is less
  sensitive to terrestrial radio-frequency interference and is narrow
  enough to resolve out extended structure in the sky, improving
  sensitivity to some pulsars (e.g the Crab pulsar). However, the
  narrow beam also has the disadvantage of a small field of view when
  the array is used for pulsar surveys.

\clearpage

\clearpage

\section{PuMa II design}

\subsection{Overview}

  Motivated by the increased bandwidth of the upgraded WSRT, we began
  searching for a new pulsar machine concept. It was realized early in
  2003 that the technology at that time had the potential to support
  the large data rates ($\approx 800 MB/s$) required to sample and
  store the entire WSRT bandwidth. It was also clear that a computer
  cluster would be needed to record and process this large data rate,
  with each computer node supporting $\approx 100MB/s$. A technology
  survey was carried out, and a single node prototype was built using
  a fast computer based on the Supermicro X5DAE
  motherboard\footnote{http://www.supermicro.com/products/motherboard/Xeon/E7505/X5DAE.cfm}. 
  The X5DAE was chosen based on the requirement
  of two independent Peripheral Component Interconnect (PCI) buses to
  support the high data rate. This machine proved to be adequate after
  a careful choice of XFS\footnote{XFS Filesystem from Silicon
  Graphics Inc. For more details, see
  http://oss.sgi.com/projects/xfs/}, a high performance filesystem, a
  fast DMA\footnote{High throughput Direct Memory Access (DMA)
  card. See http://www.edt.com/pcicda.} card and a high speed disk
  pack based on 3Ware\footnote{High performance disk access cards from
  3Ware Inc. See http://www.3ware.com/products/serial\_ata9000.asp}
  Redundant Array of Independent Disks (RAID) card. In the prototype,
  the storage medium was eight parallel IDE devices tied in a RAID0
  configuration to support high throughput. The XFS filesystem was
  configured with the real time subvolume option, to support low
  latency, high speed data writes to the disk surface. With the
  real time subvolume option, the block buffering in Linux kernel
  2.4.25 is bypassed, permitting fine grained control of the data
  writes to the disk. This configuration allowed a maximum throughput
  of $\approx 100 MB/s$. The prototype design was based on an analogue
  input to the system. An 8-bit dual converter sampled the analog
  signal at 40 MHz. Later in the prototype testing stage, it was
  realized that the digital outputs of the TAAM can be used. Therefore
  an interface card, the PuMa II interface Card (PiC), was designed to
  accept the digital input (see next subsection for details).

\clearpage

\begin{figure}[htbp]
  \epsscale{.98}
  \plotone{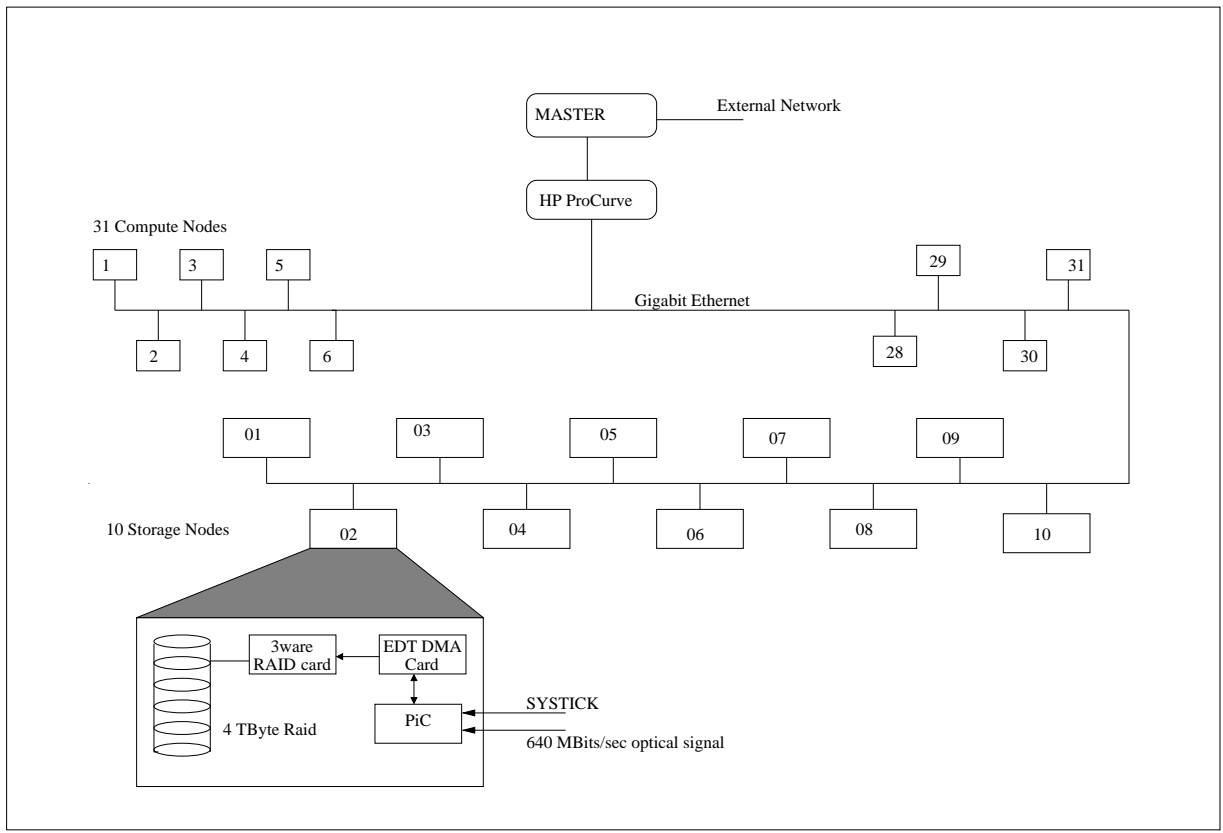}
  \caption{ A simplified block diagram of the Pulsar Machine II (PuMa
  II). The inset shows the details of a storage node. The PuMa II
  Interface Card (PiC) is the only custom hardware in the system.}
  \label{fig2}
\end{figure}

\clearpage

  With the knowledge acquired in the prototyping stage, the final
  baseband recording system was designed with 8 storage nodes (plus 2
  spares) and 32 nodes for computing. Two nodes each equipped with a
  8-tape juke-box were added later for the archiving of reduced
  data. The current and final configuration of PuMa II is a cluster of
  44 computers connected by gigabit Ethernet and is shown in Figure
  \ref{fig2}. Each node consists of a S2882\footnote{Server grade
  motherboards from Tyan Inc. see
  http://www.tyan.com/products/html/thunderk8spro.html } motherboard,
  with dual-Opteron processors clocked at 2.0 GHz. A HP Procurve
  gigabit switch is used for the cluster network. The storage nodes
  are equipped with an EDT DMA card, and 3Ware RAID card, a PiC and
  4TB of disk space. The separation of acquisition and processing
  aspects in the system described above easily meets the high speed
  sustained recording speeds of 80 MB/s per node amounting to a total
  throughput of 640 MB/s for the whole system while offering near
  real time coherent dedispersion for a range of pulsars.

\subsection{PuMa II interface Card}

\begin{figure}[htbp]
  \epsscale{.94}
  \plotone{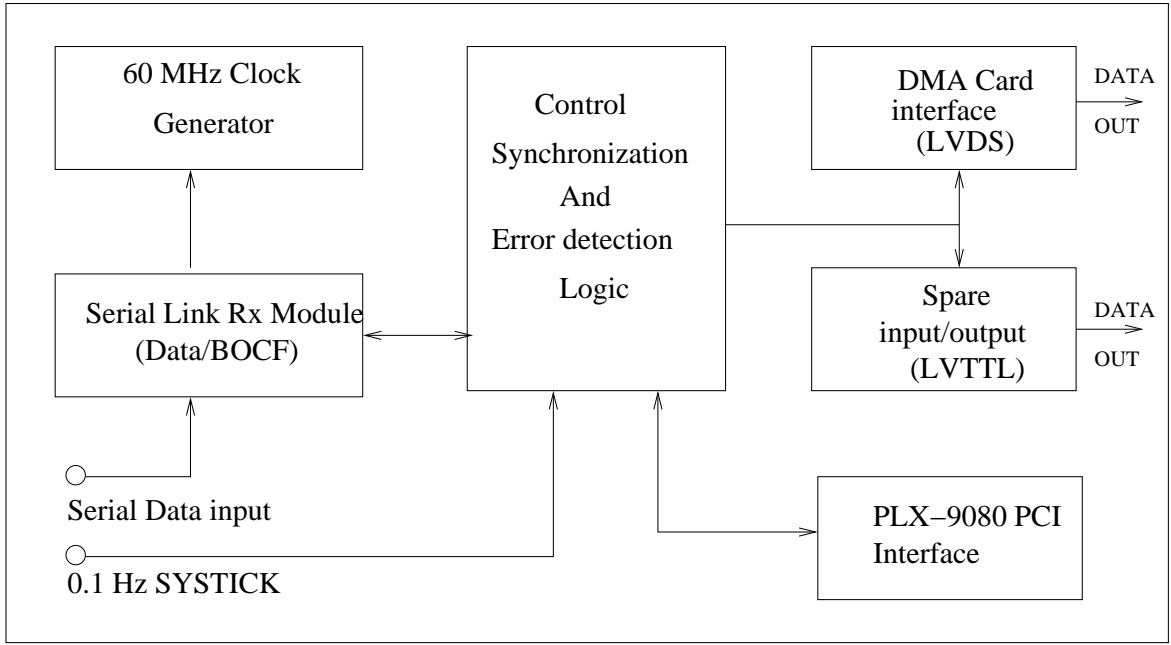}
  \caption{ A block diagram of the PuMa II Interface Card. The DMA
  interface is implemented using LVDS drivers and 80-pin high density
  connector. The 0.1 Hz observatory reference is supplied via a
  separate connector, and the serial data stream is connected using
  50$\mu$m multi-mode fiber. After the card is plugged in to the PCI
  slot of the motherboard, all PiC connectors are accessible from the
  computer's back panel. See main text for more details.}
  \label{fig3}
\end{figure}

 As described above, the optical nature of the subsystem communication
 links required that a custom interface card be built. The PuMa II
 Interface Card (PiC), was designed as an electronic card that
 integrates well in the PuMa II cluster. The card is compatible with
 the PCI bus, a standard connection system in computer
 motherboards. Figure \ref{fig3} shows a block diagram of the PuMa II
 Interface card. The card is realized on an 8-layer, short sized
 printed circuit board (PCB). The PCB was designed to allow high clock
 rates, up to 200 MHz and hence differential signal lines are used
 extensively. The low voltage differential signaling (LVDS) technique
 permits much larger clock rates (up to $\approx$ 650 MHz), while
 preserving signal integrity within the PCB. The integrated circuit,
 PLX 9080 provides the PCI bus interface on this card. A Field
 Programmable Gate Array (FPGA) provides board control logic, data
 synchronization, error detection and monitor logic. The digital
 optical signal from TAAM is converted to electric signals in the
 serial link receiver unit. This is a piggy back module used in
 various subsystems of the WSRT and is designed around an Intel
 TXN31011 optical transceiver chip. The output of the receiver module
 is 2$\times$8-bit digital data that are synchronized in the FPGA using
 the 0.1 Hz system tick (SYSTICK, derived from the observatory's MASER
 reference). The data from the FPGA are converted to LVDS using
 differential driver chips and sent to the EDT DMA card via a short
 80-core shielded, twisted pair cable.
 
\clearpage

\clearpage

 The data in the PiC are synchronized in two stages. The first level
 of coarse synchronization is done when the upper layer software
 requests a recording to be started. The start command is honored
 only if the command arrives before the 9th second of a 10 second
 synchronization boundary. On receiving this command, the PiC is
 armed, and waits for the next possible 10-second hardware trigger,
 the SYSTICK, to arrive. Once the trigger arrives, the synchronization
 is done on the fine grained sync signal called Begin of Correlator
 Frame (BOCF) is delivered in the serial data stream. From this moment
 onwards, the data are allowed to flow into the FIFO in the DMA
 card. The first sample is then timestamped based on the 0.1 Hz
 hardware synchronization signal. A byte count is maintained
 throughout the observation, thus providing accurate time stamps for
 all subsequent samples recorded as data files on the disk. The data
 written are monitored by the byte count and a resynchronization
 procedure can be initiated if data loss is detected. In practice this
 is almost never done as the system reliability easily allows
 uninterrupted recording of 6 hours, which is determined by the
 largest contiguous disk volume.

\subsection{System Software}

\begin{figure}[htbp]
  \epsscale{1.08} 
  \plotone{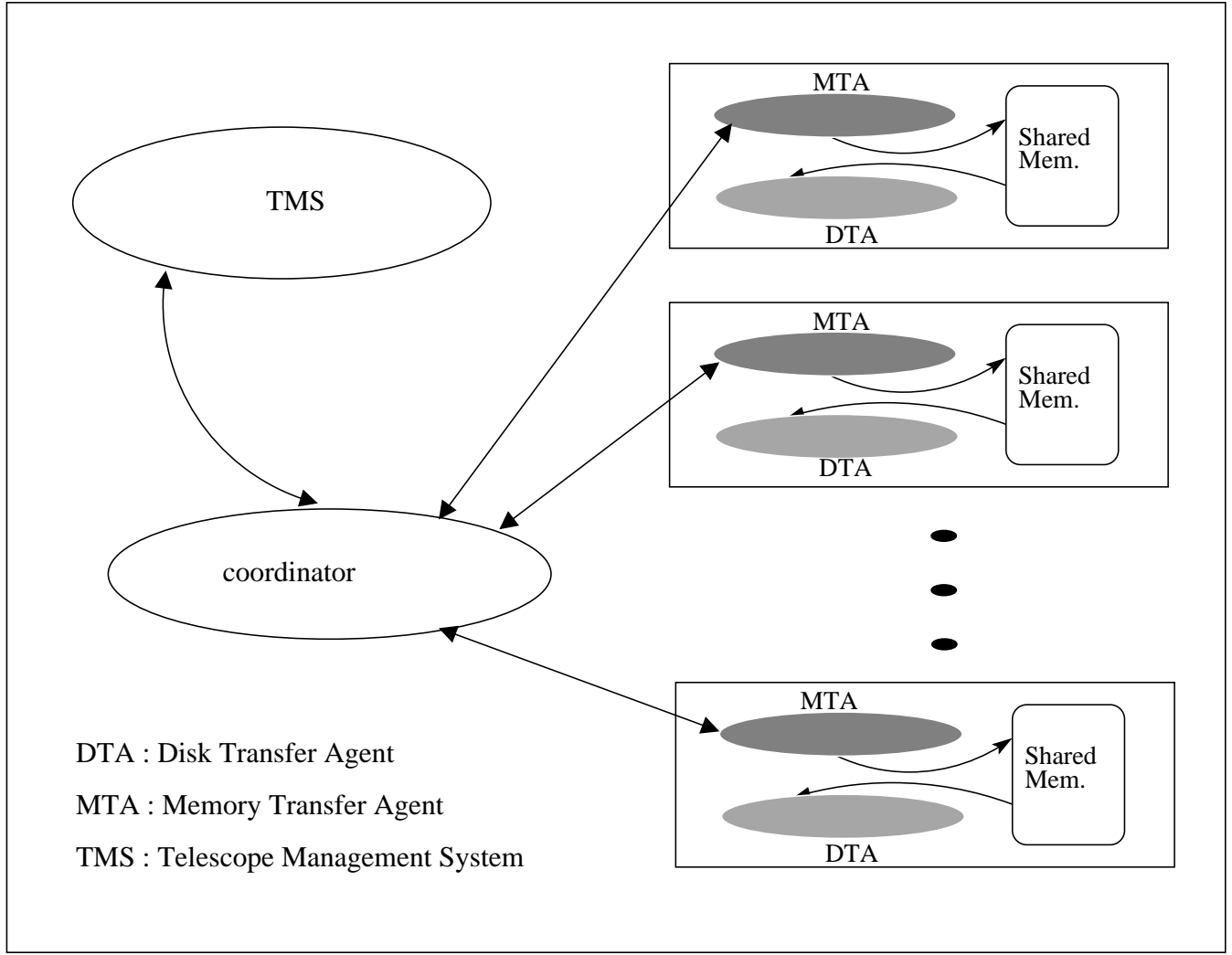}
  \caption{ Software components in the PuMa II baseband recorder. The
    software abstraction of various hardware sections are shown. Here,
    the MTA transfers data from the direct memory access hardware to
    the physical memory. The disk transfer agent writes data from the
    physical memory to the disk. The coordinator software component
    shown provides an interface to the Telescope Management System
    (TMS).}
  \label{fig4}
\end{figure}

 The software in PuMa II consists of three different aspects,
 providing complete flexibility of the instrument operation. Figure
 \ref{fig4} provides an overview of the software components relevant
 to PuMa II. Many of the components are not visible to the end user,
 as the whole system is abstracted by the top layer software. All
 software components rely on socket based communication, except for
 those within a node, which communicate via shared memory structures.
 
\clearpage

\clearpage

\subsubsection{Data acquisition software}

 The data acquisition software runs in the storage nodes, and is
 responsible for the high speed baseband recording in PuMa II. This
 module includes software to control the PiC, EDT DMA card and data
 writes to the hard disk. Two multi-threaded processes, the first for
 memory transfers and the second for disk transfers, run
 concurrently. The former controls and transfers data from the DMA
 card to a buffer in main memory, while the latter flushes data out
 from main memory to the hard disk. Both processes update their
 progress in a shared memory area. The buffer in main memory is
 organized as a collection of shared memory segments of 800 MB total
 size and can cope with disk latencies up to 10 seconds. A collection
 of 16 kernel buffers, each of 4MB size arranged as a ring buffer, are
 reserved for the DMA transfers. These DMA transfers are the lowest
 level of data transfer and they are handled by the Linux kernel
 module of the DMA card. The memory transfer agent copies this data to
 the buffer in the main memory. Since the ADC units of the WSRT always
 sample at a 40 MHz rate even if the input band is selected to be less
 than 20 MHz, decimation of data can be done at this point to reduce
 the amount of data written to the disks. The decimation factor can be
 1,2,4,8 or 16 corresponding to input bands of 20, 10, 5, 2.5 or 1.25
 MHz respectively. This feature is useful when observations are done
 at low sky frequencies, where only relatively interference free
 regions of a band are to be recorded. The disk transfer agent waits
 on a filled buffer, and it is then written to the disk when the
 buffer is signaled full by the memory transfer agent.

 One thread of the memory transfer process listens to a Unix socket,
 allowing control from upper layer software. Using this interface a
 recording can be started or stopped. Other status information is also
 exchanged via this interface. The two processes (disk and memory
 transfer agents) run in all storage nodes, and a third software
 component communicates to all eight storage nodes. This approach
 encapsulates the PuMa II baseband system as a single instrument by
 the telescope software.

 The disk transfer software can be replaced by a network transfer
 program (still in development). This distributes data directly to the
 compute nodes in the cluster, allowing network recording or real time
 distributed processing of baseband data. In the network storage mode,
 each storage node needs to be assigned at least $N_n$ target compute
 nodes and is determined by,

\begin{equation}
 N_n =\frac{80}{R_{n}},
\end{equation}

 where, $R_{n}$ is the disk throughput rate in MB/s of the target
 compute node. For the compute nodes $R_n$ is $\approx 30 MB/s$,
 giving $N_n = 3$. The assumption here is that network speed is equal
 to or better than 80 MB/s, which is true for the gigabit network used
 in the PuMa II cluster. In the compute nodes, the disk transfer
 process can be used to write the data to the disk. The processing
 software can be used if the data can be processed in real time.

 \subsubsection{Processing software}

 The main objective of pulsar signal processing is the removal of
 dispersion suffered by pulsar signals as they propagate through the
 ISM. The ISM can be seen as a tenuous ionized plasma that introduces
 a frequency dependent delay to the radio waves traveling in
 them. Dispersion smears the signal across the observing band, and
 limits time resolution if uncorrected. The effect can be modeled with
 a filter that phase shifts the signal depending on its
 frequency. Following \citet{hr75}, the transfer function of the
 filter can be expressed as:
  
\begin{equation}
H(f_0 + f) = \exp{\bigg (i \frac{2 \pi D f^2}{{f_0}^2(f_0 + f)} \bigg)},
\label{eqn1}
\end{equation}

 where $f_0$ is the mid frequency of the observing band, $f$ is
 observed frequency band, with $f\ll f_0$ and D is the dispersion
 constant defined by, 

$\displaystyle D=\frac{DM}{2.41 \times 10^{-10}}$, and 
$\displaystyle DM=\int^{d}_{0}n_e dl$,

 where $DM$ is the dispersion measure in pc cm$^{-3}$, $d$ is distance
 to the pulsar, and $n_e$ is the integrated electron density along the
 line of sight to the pulsar. The dispersive effect of the ISM can be
 undone numerically, if the pulsar signal is recorded as baseband
 voltages. This is done by convolving the data with the inverse of
 \ref{eqn1} and is called coherent dedispersion or pre-detection
 dedispersion.

 The coherent dedispersion technique is computationally very intensive
 and the resulting time resolution is not required for most pulsar
 studies. Therefore a combination of synthetic filterbanks and
 coherent dedispersion can be used to remove dispersion in a
 computationally efficient way. This method is called the coherent
 filterbank. The synthetic filterbank is formed in software by
 computing the Discrete Fourier Transform and can be efficiently
 calculated using the Fast Fourier Transform (FFT) algorithm
 \citep{vai92}. An implementation of the coherent filterbank is
 described by \citet{jcpu97}, where an N-channel synthetic filterbank
 is formed first by segmenting the real data in to sequences of length
 2N and FFTed to give N complex points. The first point from
 successive transformed sequences form the time series from the 1st
 filter channel, the second point is the time series from the 2nd
 channel and so on. In the second step, the time series from each of
 these synthetic filterbank channels are dedispersed by convolving
 with the dedispersion response function. This removes the dispersion
 introduced by the ISM completely in the filter channels. As a last
 step, the correction of dispersion across the channels is done by
 introducing a time shift in the filter channel, calculated from the
 center frequency of the channel and the $DM$ of the pulsar.

 The coherent filterbank method described above suffers from spectral
 leakage, where the power from adjacent filter channels have only 13dB
 of suppression, or about 20\% of the power from adjacent channels
 leaks into any given channel. To reduce spectral leakage, one can
 take longer FFTs or use a windowing function on the data before
 taking the FFT. The former method \citep{van03a} is used in our
 software. In our implementation, as a first step, a large K-point
 forward FFT is computed, where $K=N \times N_c$ and $N_c$ is the
 number of filter channels required. In the second step, each N-point
 segment is multiplied by the dedispersion response function tuned to
 center frequency of the channel. This is then followed by $N_c$
 inverse N-point FFTs, giving an N-point time series in each of $N_c$
 channels. Using the standard TEMPO timing solution, data is folded at
 the pulse period in each channel, resulting in $N_c$ average pulse
 profiles. If desired, single pulse outputs are written to the
 disk. The method described here is available as an open source
 package, DSP for Pulsars,
 DSPSR\footnote{http://dspsr.sourceforge.net/}. The
 PSRCHIVE\footnote{http://psrchive.sourceforge.net/} \citep{hvm04}
 utilities provide additional analysis and viewing capabilities. Using
 PSRCHIVE utilities, full Stokes parameters can be formed, reduced
 data can be converted to other formats, data affected by interference
 removed from selected channels, and reduced data can be viewed.

 For single pulse work, the DSPSR reduction software currently allows
 data reduction up to a maximum time resolution of 25ns, which is
 permitted by the sampling clock of the pulsar signal and the WSRT
 Tied Array Interface. Data from the maximum possible eight 20MHz
 subbands can be combined in software as discussed in \citet{spb+04},
 resulting in a maximum time resolution of 3.125ns. This software is
 still under development.

 The compute cluster in PuMa II is used to process data using this
 software. The cluster runs an open source version of Grid Engine
 utility from Sun Microsystems Inc., with which the processing tasks
 are submitted to a job queue. The data are read from the storage
 cluster, and reduced in the compute cluster. The results are copied
 to the archival nodes, and written to magnetic tapes using another
 utility.
 
\subsubsection{Telescope software}

The WSRT uses a large software application, the Telescope Management
System (TMS), to control telescope tracking, feeds in the frontend,
and all of the backend hardware. PuMa II has a software interface to
TMS giving full control over data acquisition. This is done in two
steps.

\begin{itemize}
\item prepare a specification file before observation
\item specify parameters and start an observation using TMS.
\end{itemize}

In step 1, the pulsar astronomer specifies details on how the data have
to be recorded which includes the frequency band to be recorded, the
down sampling rate and the data disk. This information is stored in a
text file. This will be expanded in the future to include processing
details like number of frequency channels needed, folding options and
single pulse dumps. The second step of an observation is the
specification of an observation using TMS. When an observation is
fully specified, TMS steers all the telescopes, chooses the right feed
in the frontend, sets the local oscillator in the frontend for the
proper sky frequency, and configures the backend hardware to add
signals. The correlator backend is enabled by default, and computes
cross correlation products across all possible baselines in the array
for all observations carried out at the WSRT. If a pulsar observation
was requested, the name of the file specified in step 1 is sent as a
parameter to PuMa II. A sub-program of TMS communicates with PuMa II
and monitors it continuously. This combination of TMS and the control
software in PuMa II makes pulsar observations very user friendly. An
additional software component in PuMa II will be added in the future
to allow automatic reduction of data.

\section{Observations}

PuMa II obtained first light on the pulsar PSR B0329+54 in April 2004,
with a 20 MHz band. The instrument has undergone several refinements
since then to support 8$\times$20 MHz dual polarization operation and
to allow full control of PuMa II by the telescope management
software. Some examples of the observations below gives a flavour of
the new instrument's capabilities in combination with the WSRT. The
data were recorded on PuMa II and processed using the coherent filter
bank software and all four polarization products were formed. The post
processing analysis was done using the PSRCHIVE utilities.

\subsection{PSR B1937+21}

\begin{figure}[htbp]
  \includegraphics[scale=0.45,angle=-90]{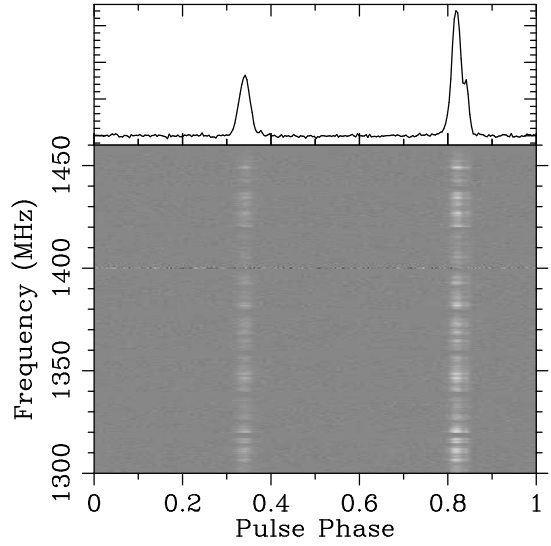}
  \caption{ An image of the dedispersed pulse stack, with phase bins
 in abscissa and frequency in ordinate. The equidistant horizontal
 bands seen at 1320,1340,1360,1380,1400,1420 and 1440 MHz are due to
 the roll off in the bandpasses of the eight 20 MHz video filters. The
 feature seen at 1400 MHz is due to interference. The upper panel
 shows the average pulse profile for the 20-minute observation.}
  \label{fig5}
\end{figure}

PSR B1937+21, the original millisecond pulsar, is the second fastest
spinning pulsar, with a period of 1.55 msec \citep{bkh+82}. The pulsar
is an isolated millisecond pulsar, with a low spin down rate and is
known to emit giant pulses \citep{cst+96}. This pulsar is observed
regularly at the WSRT in several frequency bands as a part of the
European Pulsar Timing Array. As an illustration of PuMa II's
capabilities, we present a 21cm band observation on 10 Nov 2006 at a
sky frequency of 1380 MHz with 160 MHz bandwidth in dual linear
polarizations. The pulsar was observed for 20 minutes and the pulsar
signal was baseband recorded as 8$\times$20 MHz subbands. The recorded
data were coherently dedispersed in the compute cluster of PuMa II,
using a 64-channel filterbank, resulting in a time resolution of
$\approx$ 6 $\mu$s. The data were then folded to generate an average
pulse profile every 10 seconds.  The baseband recording generates an
800 MB file every 10 seconds, and the 20-minute observation resulted
in 120 files amounting 96 GB per node or a total of 768 GB for all
eight nodes. The 10-second average pulse profiles are integrated in
time, and then combined in frequency using PSRCHIVE utilities. Figure
\ref{fig5} shows the results from processed and reduced data. The
typical error in an individual time of arrival (TOA) measurement of
PSR B1937+21 with PuMa II is 60 ns. The best TOA error that could be
measured with the PuMa I system, for an equivalent observation
duration, was 150 ns and the typical error was $\approx$280 ns. We
have not yet been observing with PuMa II sufficiently long to be able
to do a clear comparison of the long term timing solution for any pulsar.

\clearpage

\clearpage

\subsection{The Crab Pulsar Giant Pulses}

 The Crab Pulsar emits broadband, narrow and very bright pulses, known
 as giant pulses (GP). The pulsar was discovered through the GP
 emission \citep{sr1968}. It has also been found that a significant
 contribution to the average radio pulse profile comes from the giant
 pulse emission \citep{psk+06}. We observed the Crab Pulsar, B0531+21
 at 1400 MHz for 6 hours on 11 October 2004. The results from the
 analysis of the data will be published elsewhere.

\clearpage

\begin{figure}[htbp]
  \includegraphics[scale=0.62,angle=-90]{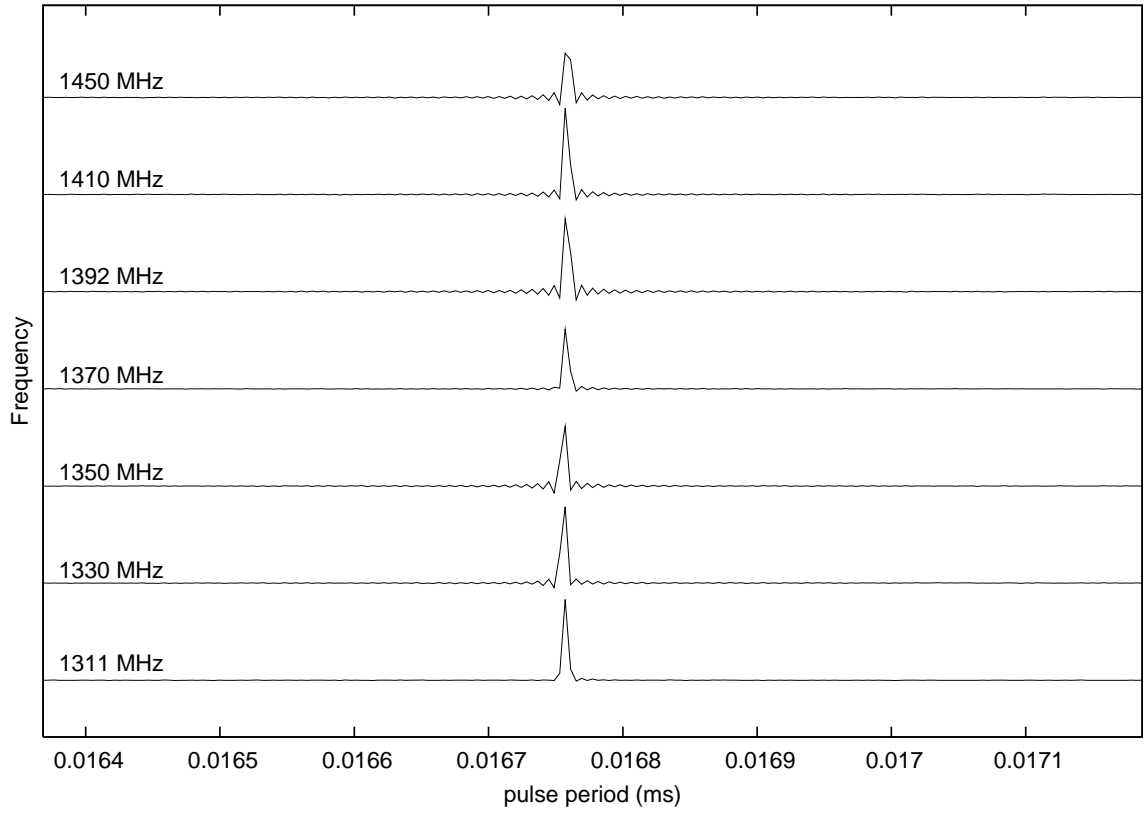}
  \caption{Plot of a broadband giant pulse detected in seven of eight
  observed bands. The 20 MHz band at 1430 MHz was not recorded due to
  hardware failure. Bands at 1311 MHz and 1392 MHz are non-uniformly
  offset to circumvent radio frequency interference. The pulses are
  dedispersed to the centre frequency of each band. The time
  resolution is 4.2 $\mu$s. The ringing effect seen in the baseline
  around the pulse is an artifact of software processing.}
  \label{fig7}
\end{figure}

\clearpage

 The data were coherently dedispersed based on the Crab Pulsar
 monthly ephemeris\footnote{http://www.jb.man.ac.uk/$\sim$pulsar/crab.html}
 maintained by Jodrell Bank Observatory. The data were processed in
 the PuMa II cluster using the coherently dedispersed single pulse
 dump mode with a 32-channel filterbank. The four polarization
 products are written to the disk per pulse period with a final time
 resolution of 4.2 $\mu$s. Even though giant pulses from the Crab are
 can be as narrow as 0.4 ns \citep{he07}, at 1380 MHz, scattering
 broadening limits time resolution to $\approx$ 4$\mu$s
 \citep{sbh+99}. In addition to this fast dump, an average pulse
 profile was formed every 10 seconds by folding the data at the pulse
 period, after dedispersion. A total of 7500 simultaneous giant pulse
 events were detected in all observed subbands. A 7$\sigma$ detection
 threshold was used to mark giant pulses. Figure \ref{fig7} shows one
 of these giant pulses that occurred simultaneously in all seven
 bands. The pulse shown at 1410 MHz has a peak S/N of 870.

\subsection{PSR B1133+16}

\begin{figure}[htbp]
  \includegraphics[scale=0.38,angle=-90]{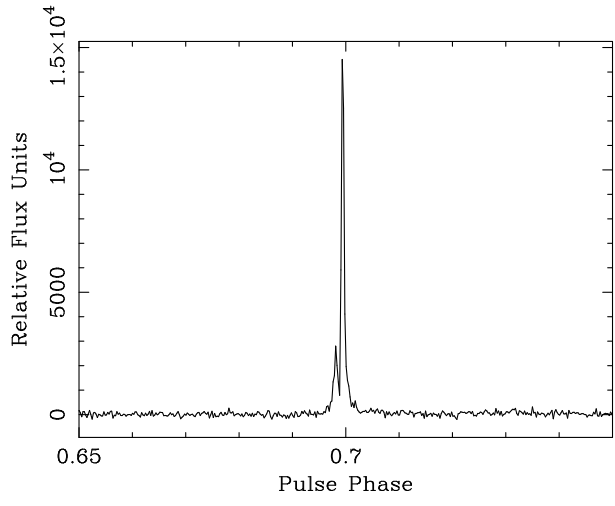} \caption{A giant
  pulse from PSR B1133+16 detected in eight frequency bands centered
  on 150 MHz. The pulse has a flux of approximately 2000 Jy and a
  width of 1 ms.}  \label{fig8}
\end{figure}

PSR B1133+16 is a nearby pulsar with a period of 1.18s and a DM of
4.864\footnote{Pulsar catalog: http://www.atnf.csiro.au/research/pulsar/psrcat}
\citep{hlk+04,mhth05}. Giant pulse like events were reported from this
pulsar \citep{kkg+03} at 4.85 GHz. PSR B1133+16 was observed at the
WSRT on 11 May 2007, at eight 2.5 MHz relatively interference free
frequency bands centred at 116.75, 130, 139.75,142.25,147.5,156,163.5
and 173.75 MHz. With a 64 channel filterbank and coherent
dedispersion, the reduced data resulted in a final time resolution of
256 $\mu$s. We have seen giant pulse like features in our data. The
narrowest pulse is 1ms wide with an intensity of 2000 Jy and the pulse
is shown in Figure \ref{fig8}. This is the first detection of giant
pulses at this low frequency and the detection of these narrow
features at low frequencies would be a formidable task with an analog
filterbank which would need to have a large number of narrow filters
to correct for dispersion effectively. The utility of software
coherent dedispersion aided by baseband recording is evident in this
case.

\clearpage

\clearpage

\subsection{PSR J1713+0747}

\begin{figure}[htbp]
  \includegraphics[scale=0.38,angle=-90]{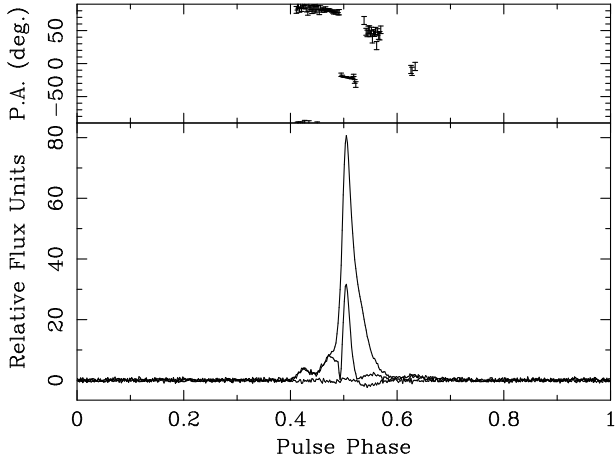}
  \caption{A polarization plot of PSR J1713+0747. The top panel shows
  position angle swing across the pulse period. Total intensity, total
  linear and circular polarizations are in the lower panel.}
  \label{fig9}
\end{figure}

The baseband sampling nature of PuMa II means that polarization data
is obtained routinely. An example of this is shown in Figure
\ref{fig9}. The millisecond pulsar PSR J1713+0747 was regularly
observed at the WSRT with PuMa I, and we use the ephemeris from these
observations to reduce the data from PuMa II. Pulsar PSR J1713+0747 was
observed with the WSRT on 2 December 2006 at 1380 MHz and 160 MHz
bandwidth. The signal was recorded in dual polarization 8$\times$20MHz
basebands on PuMa II. The data were coherently dedispersed and folded
using a 64-channel (in each 20 MHz band) filterbank. All four
polarization products were formed. Comparison with the most recently
published polarization profile of this source by \citet{ovhb04} shows an
increased level of linearly polarized flux in the PuMa II data.  This
can be understood by considering the two-bit quantization artifact
known as scattered power \citep{ja98}, which adds $\approx 12\%$
unpolarized flux to the digitized signal.  To systematically alter the
integrated profile, the scattered power must vary as a function of
pulse phase, which is true when the dispersion smearing across the
digitized band is less than the pulsar's spin period.  For the CPSR-II
observations of PSR J1713+0747 presented by \citet{ovhb04} (a 64 MHz
band centred at 1373 MHz), the dispersion smearing is approximately
3.3 ms, which is less than the spin period of $\approx$ 4.57 ms.
Therefore, the \citet{ovhb04} result has been depolarized during
two-bit digitization.

\clearpage

\clearpage

The signals from the 14 telescopes of the WSRT are also quantized
using two bits; however, because the scattered power does not add
coherently, there is a $\sqrt14$ reduction of quantization noise in
the signal output by the TAAM. The difference between $\approx12\%$
depolarization of CPSR-II data and $\approx3\%$ depolarization of
PuMa II data can account for the $\approx10\%$ increase in the degree
of linear polarization observed in Figure \ref{fig9}.

\section{Similar Instruments}
 
 Most current generation pulsar machines are fully digital, except for
 the very first input stages. The cost of computation has decreased
 continuously permitting widespread use of coherent dedispersion. High
 time resolution can only be achieved if large bandwidths are
 coherently dedispersed and this needs raw voltages from the
 telescope, recorded as baseband signal on disk or magnetic
 tape. However this method cannot be used on large bandwidth,
 multi-bit data for long recording duration due to the prohibitive
 date rates and volume. It is therefore useful to break down large
 bands into smaller subbands, and then apply coherent
 dedispersion. The resulting data rates are more manageable. This has
 an added advantage of efficient data reduction by distributing the
 data in a computer cluster environment. Some of these systems are
 compared with PuMa II.
 
 \citet{jcpu97} built recorders based on magnetic media that provided
 up to 50 MHz bandwidth.\citet{van03a} describes baseband recorders
 that allowed up to 128 MHz bandwidth. However, all these systems were
 2-bit recorders, requiring extensive quantization
 correction.\citet{vkv02} designed PuMa I, which provided limited
 baseband recording capability, and a flexible digital
 filterbank. This was based on the digital signal processor
 technology. Instruments like
 ASP\footnote{http://astro.berkeley.edu/$\sim$dbacker/asp.html} offered up
 to 64 MHz of 4-bit baseband recording. Another coherent dedispersion
 instrument supporting up to a maximum of 100 MHz bandwidth is
 COBRA\footnote{http://www.jb.man.ac.uk/$\sim$pulsar/cobra/}, where the
 signal is sampled using 8-bit and 10$\times$10MHz subbands. In PuMa
 II the signal is sampled and recorded as 8-bit values supporting a
 total bandwidth of 160 MHz. Coherent dedispersion is the default mode
 of operation in PuMa II. Other very wide band ($ >$600 MHz) instruments
 \citep[Spigot:][]{kel+05} are based on autocorrelation spectrometers
 allowing incoherent dedispersion. Such wide band systems are,
 however, more common in single dish telescopes, due to complexities
 in wide band beam forming in a synthesis array.

 Hardware based techniques is another possibility to process large
 volumes of data efficiently. Such systems were built in the past,
 based on discrete integrated circuits with limitations in bandwidth
 that can dedispersed, the observing frequency and the DM of the
 pulsar \citep{bdz+97}. Moreover in such systems, the data were
 averaged immediately which was not amenable to fast dump modes
 limiting the detection of single pulses. The progress in FPGA
 technology holds some promise in this direction, and is currently
 being explored (private communication, Scott N. Ransom, NRAO). This
 instrument\footnote{https://wikio.nrao.edu/bin/view/CICADA/NGNPP}(under
 development), will support up to 800 MHz of bandwidth with the signal
 sampled at 8-bit resolution. FPGA based coherent dedispersion will be
 done on smaller subbands and a limited range of pulsar dispersion
 measures.
  
 \subsection{Comparison with PuMa I}
 
 PuMa I has been the work horse in all pulsar observations carried out
 at the WSRT. The instrument records a maximum of 10 MHz dual
 polarization signals in the 2-bit baseband recording mode, or 80 MHz
 when operating in incoherent filter bank mode. PuMa I can also
 operate as a 8-bit baseband recorder, if the bandwidth is reduced to
 2.5 MHz. PuMa II operates only in baseband recording mode, and
 supports 8-bit sampling and a 160 MHz of bandwidth. Since the
 sensitivity to pulsar signals is $\propto 1/\sqrt{\Delta f.t}$, where
 $\Delta f$ is the bandwidth used and $t$ is the integration time,
 the large bandwidth in PuMa II has a clear advantage in terms of time
 resolution achievable and sensitivity. The sensitivity improvement
 in PuMa II is at least $\sqrt 2$, when PuMa I operates in the 80MHz
 filterbank mode. The improvement is a factor of 4, when PuMa I is
 used in the 10MHz baseband recording mode.  In practice, PuMa II is
 more flexible in operation, whereas the design of PuMa I requires
 specialist knowledge. PuMa II is being used successfully in all
 millisecond pulsar observations with an excellent time resolution. As
 discussed earlier, the timing of PSR B1937+21 shows at least a four
 fold decrease in time of arrival uncertainty using PuMa II. For
 pulsars like PSR J0034-0534, the brightest millisecond pulsar at low
 radio frequencies, this improvement is even more pronounced as the
 advantages of coherent dedispersion are more apparent.

 The 8-bit operation of PuMa II has an advantage over the 2-bit
 baseband mode of PuMa I. The S/N of the digitized pulsar signal is a
 function of the number bits used to represent the signal
 \citep{kv01}, and hence PuMa II offers a better dynamic range to the
 input signal. The improved dynamic range is important when
 radio-frequency interference (RFI) rejection algorithms are used to
 mitigate interference. RFI is detrimental to all radio astronomy
 experiments, since they add or remove information from the radio
 signals of astronomical origin. \citet{sst+00} describes a method to
 combat RFI, where the signal is cleaned by examining the data in both
 time and Fourier domains to eliminate broadband and narrow band
 interferences respectively. Such techniques are effective when the
 signal has a high dynamic range resulting from the multi-bit
 sampling.
 
 The design of PuMa I and its interface to the WSRT involves sampling
 of analog voltages twice in the signal path. This introduces
 quantization noise, making PuMa I less sensitive, while PuMa II has a
 single quantization step in the signal chain improving the noise
 performance.
 
 \section{Conclusion}

 We have built and installed a new pulsar machine, PuMa II at the
 WSRT. The basic mode of operation in PuMa II is the distributed baseband
 recording mode, where a 160-MHz dual-polarization analog band is
 digitized as 2$\times$8$\times$20 MHz subbands with 8-bit resolution
 and stored on disk attached to storage nodes. The data are processed
 in a separate computer cluster. This is done in PuMa II by the
 separation of a single cluster into three subclusters: a 10-node
 storage cluster with large disks for distributed recording, a 32-node
 compute cluster for data processing and two nodes equipped with high
 density tape drives for archiving reduced data. The separation allows
 a large throughput of 640 MB/s to the disks in baseband recording
 mode, while data are farmed out to the compute cluster for further
 processing. This results in a near real time data reduction for most
 pulsars observed at the WSRT. A total disk space of 32 TB
 distributed in 8 storage nodes allows 12 hours of continuous
 recording. Software based data processing makes PuMa II a flexible
 instrument, addressing a wide variety of signal processing tasks. For
 pulsar science, a software based coherent dedispersion method offers
 a routine time resolution up to 50ns. The instrument has excellent
 performance in terms of time resolution and flexibility. It also
 illustrates that large bandwidth baseband recording is possible, and
 future technology will offer even larger bandwidths. The software
 based processing offers full flexibility in processing the data,
 including RFI excision, which is becoming more important as the
 spectrum is crowded with commercial operators. The modular design of
 the instrument allows straight forward upgrades, for either larger
 bandwidth or longer storage. Possibilities exist to make the
 instrument even more flexible, with additional hardware under
 consideration, for e.g FPGA based hardware polyphase
 filterbanks. Combined with data averaging, this would reduce data
 rate, and allow longer integration times.
 
\section{Acknowledgments}

\bibliographystyle{apj}

\begin{thebibliography}{32}
\expandafter\ifx\csname natexlab\endcsname\relax\def\natexlab#1{#1}\fi

\bibitem[{{Baars} \& {Hooghoudt}(1974)}]{bh74}
{Baars}, J.~W.~M. \& {Hooghoudt}, B.~G. 1974, \aap, 31, 323

\bibitem[{Backer {et~al.}(1997)Backer, Dexter, Zepka, D., Wertheimer, Ray, \&
  Foster}]{bdz+97}
Backer, D.~C., Dexter, M.~R., Zepka, A., D., N., Wertheimer, D.~J., Ray, P.~S.,
  \& Foster, R.~S. 1997, \pasp, 109, 61

\bibitem[{Backer {et~al.}(1982)Backer, Kulkarni, Heiles, Davis, \&
  Goss}]{bkh+82}
Backer, D.~C., Kulkarni, S.~R., Heiles, C., Davis, M.~M., \& Goss, W.~M. 1982,
  \nat, 300, 615

\bibitem[{{Cognard} {et~al.}(1996){Cognard}, {Shrauner}, {Taylor}, \&
  {Thorsett}}]{cst+96}
{Cognard}, I., {Shrauner}, J.~A., {Taylor}, J.~H., \& {Thorsett}, S.~E. 1996,
  Astrophys. J., Lett., 457, 81

\bibitem[{Demorest(2007)}]{dem07}
Demorest, P. 2007, PhD thesis, University of California, Berkeley

\bibitem[{{Hanado} {et~al.}(1995){Hanado}, {Imae}, \& {Sekido}}]{his95}
{Hanado}, Y., {Imae}, M., \& {Sekido}, M. 1995, IEEE Transactions on
  Instrumentation and Measurement, 44, 107

\bibitem[{Hankins(1971)}]{han71}
Hankins, T.~H. 1971, Astrophys. J., 169, 487

\bibitem[{{Hankins} \& {Eilek}(2007)}]{he07}
{Hankins}, T.~H. \& {Eilek}, J.~A. 2007, ArXiv e-prints, 708

\bibitem[{{Hankins} \& {Rickett}(1975)}]{hr75}
{Hankins}, T.~H. \& {Rickett}, B.~J. 1975, in Methods in Computational Physics
  Volume 14 --- Radio Astronomy (New York: Academic Press), 55--129

\bibitem[{Hankins {et~al.}(1987)Hankins, Stinebring, \& Rawley}]{hsr87}
Hankins, T.~H., Stinebring, D.~R., \& Rawley, L.~A. 1987, Astrophys. J., 315,
  149

\bibitem[{{Hessels} {et~al.}(2006){Hessels}, {Ransom}, {Stairs}, {Freire},
  {Kaspi}, \& {Camilo}}]{hrsf+06}
{Hessels}, J.~W.~T., {Ransom}, S.~M., {Stairs}, I.~H., {Freire}, P.~C.~C.,
  {Kaspi}, V.~M., \& {Camilo}, F. 2006, Science, 311, 1901

\bibitem[{Hobbs {et~al.}(2004)Hobbs, Lyne, Kramer, Martin, \& Jordan}]{hlk+04}
Hobbs, G., Lyne, A.~G., Kramer, M., Martin, C.~E., \& Jordan, C. 2004, \mnras,
  353, 1311

\bibitem[{{Hotan}(2005)}]{hot05}
{Hotan}, A.~W. 2005, in Astronomical Society of the Pacific Conference Series,
  Vol. 328, Binary Radio Pulsars, ed. F.~A. {Rasio} \& I.~H. {Stairs}, 377--+

\bibitem[{{Hotan} {et~al.}(2004){Hotan}, {van Straten}, \&
  {Manchester}}]{hvm04}
{Hotan}, A.~W., {van Straten}, W., \& {Manchester}, R.~N. 2004, Proc. Astron.
  Soc. Aust., 21, 302

\bibitem[{Jenet \& Anderson(1998)}]{ja98}
Jenet, F.~A. \& Anderson, S.~B. 1998, \pasp, 110, 1467

\bibitem[{Jenet {et~al.}(1997)Jenet, Cook, Prince, \& Unwin}]{jcpu97}
Jenet, F.~A., Cook, W.~R., Prince, T.~A., \& Unwin, S.~C. 1997, \pasp, 109, 707

\bibitem[{{Kaplan} {et~al.}(2005){Kaplan}, {Escoffier}, {Lacasse}, {O'Neil},
  {Ford}, {Ransom}, {Anderson}, {Cordes}, {Lazio}, \& {Kulkarni}}]{kel+05}
{Kaplan}, D.~L., {Escoffier}, R.~P., {Lacasse}, R.~J., {O'Neil}, K., {Ford},
  J.~M., {Ransom}, S.~M., {Anderson}, S.~B., {Cordes}, J.~M., {Lazio},
  T.~J.~W., \& {Kulkarni}, S.~R. 2005, \pasp, 117, 643

\bibitem[{{Kouwenhoven} \& {Vo{\^u}te}(2001)}]{kv01}
{Kouwenhoven}, M.~L.~A. \& {Vo{\^u}te}, J.~L.~L. 2001, Astron. Astrophys., 378,
  700

\bibitem[{{Kramer} {et~al.}(2003){Kramer}, {Karastergiou}, {Gupta}, {Johnston},
  {Bhat}, \& {Lyne}}]{kkg+03}
{Kramer}, M., {Karastergiou}, A., {Gupta}, Y., {Johnston}, S., {Bhat},
  N.~D.~R., \& {Lyne}, A.~G. 2003, Astron. Astrophys., 407, 655

\bibitem[{{Manchester} {et~al.}(2005){Manchester}, {Hobbs}, {Teoh}, \&
  {Hobbs}}]{mhth05}
{Manchester}, R.~N., {Hobbs}, G.~B., {Teoh}, A., \& {Hobbs}, M. 2005, Astron.
  J., 129, 1993

\bibitem[{Navarro(1994)}]{nav94}
Navarro, J. 1994, PhD thesis, California Institute of Technology

\bibitem[{{Ord} {et~al.}(2004){Ord}, {van Straten}, {Hotan}, \&
  {Bailes}}]{ovhb04}
{Ord}, S.~M., {van Straten}, W., {Hotan}, A.~W., \& {Bailes}, M. 2004, \mnras,
  352, 804

\bibitem[{{Popov} {et~al.}(2006){Popov}, {Soglasnov}, {Kondrat'ev}, {Kostyuk},
  {Ilyasov}, \& {Oreshko}}]{psk+06}
{Popov}, M., {Soglasnov}, V., {Kondrat'ev}, V., {Kostyuk}, S., {Ilyasov}, Y.,
  \& {Oreshko}, V. 2006, Astronomy Letters, 50, 55

\bibitem[{Sallmen {et~al.}(1999)Sallmen, Backer, Hankins, Moffett, \&
  Lundgren}]{sbh+99}
Sallmen, S., Backer, D.~C., Hankins, T.~H., Moffett, D., \& Lundgren, S. 1999,
  Astrophys. J., 517, 460

\bibitem[{{Soglasnov} {et~al.}(2004){Soglasnov}, {Popov}, {Bartel}, {Cannon},
  {Novikov}, {Kondratiev}, \& {Altunin}}]{spb+04}
{Soglasnov}, V.~A., {Popov}, M.~V., {Bartel}, N., {Cannon}, W., {Novikov},
  A.~Y., {Kondratiev}, V.~I., \& {Altunin}, V.~I. 2004, Astrophys. J., 616, 439

\bibitem[{{Staelin} \& {Reifenstein}(1968)}]{sr1968}
{Staelin}, D.~H. \& {Reifenstein}, E.~C. 1968, Science, 162, 1481

\bibitem[{Stairs {et~al.}(2000)Stairs, Splaver, Thorsett, Nice, \&
  Taylor}]{sst+00}
Stairs, I.~H., Splaver, E.~M., Thorsett, S.~E., Nice, D.~J., \& Taylor, J.~H.
  2000, \mnras, 314, 459, (astro-ph/9912272)

\bibitem[{Stinebring {et~al.}(1992)Stinebring, Kaspi, Nice, Ryba, Taylor,
  Thorsett, \& Hankins}]{skn+92}
Stinebring, D.~R., Kaspi, V.~M., Nice, D.~J., Ryba, M.~F., Taylor, J.~H.,
  Thorsett, S.~E., \& Hankins, T.~H. 1992, Rev. Sci. Instrum., 63, 3551

\bibitem[{Stinebring {et~al.}(1990)Stinebring, Ryba, Taylor, \&
  Romani}]{srtr90}
Stinebring, D.~R., Ryba, M.~F., Taylor, J.~H., \& Romani, R.~W. 1990, \prl, 65,
  285

\bibitem[{Vaidyanathan(1992)}]{vai92}
Vaidyanathan, P. 1992, Multirate Systems and Filter Banks (Prentice Hall),
  113--117

\bibitem[{van Straten(2003)}]{van03a}
van Straten, W. 2003, PhD thesis, Swinburne University of Technology

\bibitem[{{V\^oute} {et~al.}(2002){V\^oute}, {Kouwenhoven}, {van Haren},
  {Langerak}, {Stappers}, {Driesens}, {Ramachandran}, \& {Beijaard}}]{vkv02}
{V\^oute}, J.~L.~L., {Kouwenhoven}, M.~L.~A., {van Haren}, P.~C., {Langerak},
  J.~J., {Stappers}, B.~W., {Driesens}, D., {Ramachandran}, R., \& {Beijaard},
  T.~D. 2002, Astron. Astrophys., 385, 733

\end{thebibliography}

\end{document}